# Navigating the Future of Education: Educators' Insights on AI Integration and Challenges in Greece, Hungary, Latvia, Ireland and Armenia

**Evangelia Daskalaki[*], Katerina Psaroudaki, Paraskevi Fragopoulou**

*Foundation for Research and Technology - Hellas (FORTH), Institute of Computer Science,*
*N. Plastira 100, Vassilika Vouton, GR-70013 Heraklion, Crete, Greece*
*Authors' emails: {eva, kpsaroudak, fragopou}@ics.forth.gr*

### ABSTRACT

Understanding teachers' perspectives on AI in Education (AIEd) is crucial for its effective integration into the educational framework. This paper aims to explore how teachers currently use AI and how it can enhance the educational process. We conducted a cross-national study spanning Greece, Hungary, Latvia, Ireland, and Armenia, surveying 1754 educators through an online questionnaire, addressing three research questions. Our first research question examines educators' understanding of AIEd, their skepticism, and its integration within schools. Most educators report a solid understanding of AI and acknowledge its potential risks. AIEd is primarily used for educator support and engaging students. However, concerns exist about AI's impact on fostering critical thinking and exposing students to biased data. The second research question investigates student engagement with AI tools from educators' perspectives. Teachers indicate that students use AI mainly to manage their academic workload, while outside school, AI tools are primarily used for entertainment. The third research question addresses future implications of AI in education. Educators are optimistic about AI's potential to enhance educational processes, particularly through personalized learning experiences. Nonetheless, they express significant concerns about AI's impact on cultivating critical thinking and ethical issues related to potential misuse. There is a strong emphasis on the need for professional development through training seminars, workshops, and online courses to integrate AI effectively into teaching practices. Overall, the findings highlight a cautious optimism among educators regarding AI in education, alongside a clear demand for targeted professional development to address concerns and enhance skills in using AI tools.

---

[*] Corresponding author: Evangelia Daskalaki, eva@ics.forth.gr, ORCID 0000-0003-3056-1311

## 1. Introduction

Artificial Intelligence (AI) has become prevalent across all levels of education. When we refer to AI in education (AIEd), we include a wide range of systems and tools that mimic human intelligence and serve various educational purposes, like teaching assistants, course creation tools, chatbots, image and video generation, audio, research, animation, text-to-speech, presentations, etc. Understanding teachers' beliefs about AIEd is essential for its effective integration into the educational system. Consequently, numerous studies have emerged in recent years to explore this topic, highlighting both educators' excitement and concerns.

The present survey about AIEd is an overview of teachers' use of AI and examines how AI can effectively support teaching. Conducted from the Greek Safer Internet Center of FORTH (Daskalaki, et al., 2020; Christodoulaki & Fragopoulou, 2010; Kokolaki, et al.,2020; Daskalaki, et al.,2018) in collaboration with the Safer Internet Centers of Hungary[1], Ireland[2], Latvia[3] and Armenia[4], and with the support of the European Network of Safer Internet Centers, Insafe[5], it represents a significant international effort. To our knowledge, this is the first international research supported by Safer Internet Centers from five countries, to investigate the use of AI in education.

This survey aims to guide investigations into various aspects of how AI is integrated into educational settings. While most research on AIEd has focused on technological improvements, according to (Kizilcec, 2024) more research of education technology from a psychological perspective is needed in order to understand what factors shape the way educators perceive, trust, and use education technology in teaching practice, leading to our **first research question: "***What is the current status of AIEd in the five countries we focus on? How do educators perceive it, and what are their primary areas of skepticism?***"** Furthermore, we were also interested in investigating the use of AI by students examining how they interact with AI technologies, their preferences, behaviors, and the impact of these interactions on their learning experiences. This forms our **second research question**: "*What are the patterns, preferences, and impacts of student engagement with artificial intelligence tools in educational settings from the perspective of educators?*". Lastly, the potential for AIEd to revolutionize how we teach, learn, and interact with educational materials is vast. This study aims to uncover potential barriers and opportunities, providing critical insight for the effective implementation of AI education programs. We explore several crucial facets of AI's forthcoming impact in our **third research question:** "*In what ways will AI shape the educational landscape from the perspective of educators?*".

The main findings of the survey are as follows:

- **Current Understanding and Utilization:** Educators across most surveyed countries report a solid understanding of AI technology and acknowledge its potential risks. Most educators from Greece, Ireland and Armenia state that they use AI tools in the educational process. AIEd is primarily used for educator support and training according to educators from Greece, Hungary, Ireland, and Latvia. Furthermore, these tools are used for engaging students in the classroom, delving deeper into the lesson, and entertaining students according to Armenian educators. Concerning student usage, educators indicate that students mainly use AI to manage their academic workload, such as completing homework effortlessly. Beyond school, educators report that AI tools are primarily used by students for entertainment purposes for Greece, Hungary, Ireland, Latvia, and Armenia. Educators also express concerns about the impact of AI on fostering critical thinking and the exposure of students to biased data for all the aforementioned countries.

- **Risks:** Considering whether they are aware of the risks for students associated with this technology, across all countries, most educators seem to be aware of the risks and raise ethical concerns, such as accountability, transparency, and bias. Concerning specific risks, in all countries the primary risk identified is students' tendency to trust the information they find without applying critical thinking. Educators from Greece, Hungary, Ireland and Armenia also call attention to the fact that students use AI to share material (e.g., photos, videos) that has been previously tampered with AI tools, which of course encloses the issue of ethical use of AI. What's more, educators from Ireland are concerned that students are exposed to biased, incorrect or harmful content. Overall, most educators who use these systems agree that there are mechanisms in place to ensure that personal and sensitive personal data are adequately protected.

- **Benefits:** According to our survey, educators unanimously believe that one of the positive effects AI will have in the future is its potential to support educators' work. Another positive impact they foresee is its potential to facilitate personalized learning experiences for students. Furthermore, most teachers believe AI will reduce their administrative tasks, allowing them to invest more time in teaching and mentoring students.

- **Concerns and Ethical Issues:** Despite their optimism, educators across all countries express significant concerns about AI's impact on fostering critical thinking and raise ethical concerns about the potential misuse of AI and the exposure of students to biased and inappropriate information. Therefore, educators emphasize the necessity for professional development to enhance their knowledge and skills, through training seminars, workshops, and specialized online courses to effectively integrate AI into their teaching practices.

- **Greek Data Analysis:** A deeper analysis of the Greek data reveals that younger educators are more likely to be tech-savvy and have experience using AI tools. Gender also influences familiarity with technology, with male educators being more familiar with technology tools, while female educators having experimented more with AIEd tools.

Overall, the findings highlight a cautious optimism among educators regarding the integration of AI in education. There is a clear demand for targeted professional development to address their concerns and enhance their skills in utilizing AI tools effectively.

The remainder of the paper is organized as follows: In section 2. Literature Review, we examine existing research and theories relevant to the study. Section 3. Survey Methodology is devoted to the methodology of the survey, including information about 3.1 Participants and Proceduresthe participants involved in the study and the procedures followed during the survey,3.2 Questionnaire and Measures as well as a description of the questionnaire used, and the measures taken to ensure data reliability and validity. Subsequently, section 4. Results 4.1Results on current landscape of AI from educators' perspectivepresents the findings for each main research question of the survey, namely, educators' views on the current state of AI in education, 4.2Results about the future of AI in educationeducators' perspectives on the future implications of AI in the educational field, and 0

**1.1. Analysis on the Greek case data**

the future of AI in education. It also provides an analysis of data specific to the Greek context, 0

Analysis on the Greek case datadelving into the Greek case. We conclude in Section 5. Discussion and Conclusion, with a summary of key findings, discussing their implications, as well as suggestions for future research.

## 2. Literature Review

In recent years, numerous researchers have dedicated substantial effort to the study of AIEd. A review of journal articles from 2010 to 2020 (Crompton & Burke, 2022) found four main ways educators used AI to help students learn: monitoring students, managing groups, automated grading, and making data-based decisions. In group management, AI assisted teachers in forming, moderating, and facilitating groups. For students, three major benefits of AI emerged: AI tutors, enhancing student thinking, and personalized learning tailored to each student's strengths, weaknesses, preferences, and interests.

According to Akgun and Greenhow (Akgun & Greenhow, 2022), many teachers believe that introducing AI concepts at an early age can foster critical thinking and problem-solving skills. This early exposure can make children feel more comfortable with technology, becoming at the same time better equipped for their future academic and/or career pursuits.

Roll and Wylie (Roll & Wylie, 2016) suggest that AIEd can relieve teachers from the need to acquire all the necessary knowledge and information their students might require. Instead, teachers can focus on providing students with methods to find, discuss, and research information themselves, through team projects or knowledge-building activities. Teachers at present spend a lot of time grading homework and tests, which takes away valuable time from their teaching and quality time with their students (Rasul et al., 2024).

Indeed, AI tools like intelligent tutor systems, assessment systems, and educational robots can take over these repetitive tasks. This helps reduce the teachers' workload and allows them to focus more on teaching and building relationships with students (Tahiru, 2021). Furthermore, teachers see AI literacy as a way to promote equity in education. As highlighted by (Beverly Park Woolf et al., 2013) AI can personalize learning experiences, making education more effective by analyzing datasets on teaching behavior, student motivation, and social interactions. This potential for personalization is particularly appealing in elementary education, where students' learning needs can vary widely.

Further research has shown that AI technology can make teaching more efficient. Additionally, AI can provide teachers with more free time and energy for more effective communication with their students, helping them focus on subjects such as morals and skills development. It can also provide teachers with the opportunity to pay more attention to each student's overall physical and mental growth (Tanveer et al., 2020). It has been noted that teachers have transitioned from a position of providing knowledge to being a medium of facilitating student learning, with a focus on student-centered education, offering at the same time more compassionate care when it is needed.

Teachers who view Generative AI from a positive side are more likely to incorporate it into their daily teaching methods, as indicated by (Kaplan-Rakowski et al., 2023). The critical factor in this statement seems to be the age of the educators, since the younger ones, particularly those from Generation Z, are more receptive and better adapting to technological advancements (Chan & Lee, 2023).

There are also voices among them who express their concerns regarding AI literacy education. (Ally, 2019) noted that all the latest technological advances have upgraded the educational system, since they have highlighted new significant alterations in teachers' instructional methods and the way the educators view themselves as a significant part in their role of the learning process. The main concern we should be worried about is the lack of sufficient training and resources. A study surveyed K-12 teachers in Serbia aiming to find out how much they know about AI, how they use it in teaching, and what they think about it (Kuleto et al., 2022). The results showed that teachers who were more likely to use AI as part of their educational method or expressed their wish to use it in their teaching, were the ones who viewed it more positively. The above finding suggests that had teachers been provided with more training on using AI, they could turn to be more willing to incorporate it into their teaching practices. On the other hand, teachers with less training and exposure to AI technology tend to be more skeptical about using it. The latter often perceive AI literacy as an additional and more complex workload rather than an opportunity to improve their teaching methods and skills. This skepticism is further enhanced by their perception of the complexity of AI tools and the difficulty they will face in making these concepts accessible to young learners.

A scoping review was performed to analyze 16 empirical studies published between 2016 and 2022 (Su et al., 2023). This review evaluated, examined, and presented research on AI literacy in early childhood education, covering curriculum design, AI tools, teaching methods, research approaches, assessment techniques, and results. The review highlighted several challenges and opportunities related to AI literacy. Key challenges included (1) a lack of AI knowledge, skills, and confidence among teachers, (2) insufficient curriculum design, and (3) a lack of teaching guidelines. Despite these initial hurdles, AI learning has the potential to create educational opportunities and enhance young children's understanding of AI concepts, practices, and perspectives. However, there are still concerns about the fear of overreliance on AI after long-term use, which may undermine traditional teaching methods and pedagogical principles, resulting to uncertain educational outcomes. Moreover, the ethical implications of introducing AI to young children is another concern in the educational community. Teachers are concerned about the existing possibility of AI amplifying issues related to privacy, data security and digital dependency (Institute for Ethical AI in Education).

Another report (Schiel et al., 2024) explored how students in the U.S. use AI tools for school assignments and other activities, their perceptions of the cognitive and academic impacts of these tools, and their views on employing AI to write their college admissions essays. Their findings were that nearly half of the high school students surveyed reported using AI tools, with ChatGPT being the most popular. Among the 54% who hadn't used AI tools, the primary reasons were a lack of interest, distrust of the information provided, and insufficient knowledge about them. Students also used AI tools for entertainment, hobbies, and personalized recommendations. Those with higher academic performance were significantly more likely to use AI tools than those with lower performance. Nearly 74% of students believed that using AI tools would slightly improve their school performance. However, 90% had not considered using AI tools for their college admissions essays, citing concerns about the tools' current limitations in generating high-quality, personalized, original, and authentic content that reflects their skills and unique writing styles. Additionally, students felt that using AI for this purpose would be dishonest and unethical and preferred the sense of accomplishment from writing their own essays.

The integration of AI into education presents a nuanced blend of opportunities and challenges. While several teachers embrace the concept of AI becoming incorporated into teaching and learning experiences, others express concerns regarding further aspects such as its financial implications, specific ethical considerations, and potential impacts on conventional pedagogical methodologies. It is vital to fully understand these divergent perspectives if we want to build new strategies that will empower teachers to harness AI in education effectively and ethically.

## 3. Survey Methodology

### 3.1. Participants and Procedures

The survey was conducted from the members of the Greek Safer Internet Center of FORTH (Daskalaki et al., 2020; Christodoulaki & Fragopoulou, 2010; Kokolaki et al., 2020; Daskalaki et al., EL-SIC: focus on better and Safer Online Experiences for Kids, 2018) in collaboration with the Safer Internet Centers of Hungary[6], Ireland[7], Latvia[8] and Armenia[9], and with the support of the European Network of Safer Internet Centers Insafe[10].

It took place between October 2023 – March 2024, and the data of the different countries were collected anonymously via online questionnaires. The online questionnaire and its translations were published in the EUSurvey online survey management system. EUSurvey is the official online survey management tool of the European Commission. Its development started under the supervision of DIGIT11 and is currently available as open-source software under the terms of the EUPL public license. The EUSurvey system adheres to the Web Content Accessibility Guidelines (WCAG) 2.0 Level AA when selected. This setting has been activated in the questionnaires, in order to be inclusive for all.

The survey involved educators from Greece, Hungary, Ireland, Latvia, and Armenia. The online questionnaire was distributed to educators directly from the respective national Safer Internet Centers. Specifically, for Greece, the questionnaire was also communicated to the educators from the Panhellenic School Network12, the Greek national network (ISP) of the Ministry of Education which safely interconnects 16K schools of Primary and Secondary education, including educational units abroad, as well as services and entities supervised by the Ministry of Education at central and regional level.

All questions and responses in the questionnaire were provided in the national language of each country to ensure maximal level of clarity and understanding for all participants.

The participants were clearly informed throughout the study that their participation in the research was completely voluntary and anonymous. A total of 1754 educators completed the online survey, of which 1125 were Greek, 298 Hungarian, 130 Irish, 70 Latvian and 131 Armenian educators. The demographic information of the collected sample can be seen in        Table **1**.

---

**Table 1.** Table showing the demographic information of the sample.

| Measure | Participants | | N | % |
|---|---|---|---|---|
| Country | Greece | | 1125 | 64% |
| | Hungary | | 298 | 17% |
| | Ireland | | 130 | 7% |
| | Latvia | | 70 | 4% |
| | Armenia | | 131 | 7% |
| | | Total | 1754 | 100% |
| Age | Under 25 y.o. | | 13 | 0.7% |
| | 26-35 y.o. | | 156 | 8.9% |
| | 36-45 y.o. | | 460 | 26.2% |
| | 46-55 y.o. | | 689 | 39.3% |
| | Over 55 y.o. | | 436 | 24.9% |
| | | Total | 1754 | 100% |
| Participants' Gender | Female | | 1220 | 69.6% |
| | Male | | 516 | 29.4% |
| | Non-binary | | 1 | 0.1% |
| | I would rather not mention | | 17 | 1.0% |
| | | Total | 1754 | 100% |
| Participants' teaching experience in years | 1-4 y. | | 128 | 7.3% |
| | 5-10 y. | | 187 | 10.7% |
| | 11-20 y. | | 525 | 29.9% |
| | 21-30 y. | | 632 | 36.0% |
| | 31+ y. | | 282 | 16.1% |
| | | Total | 1754 | 100% |
| Place of Residence | Village | | 196 | 11.2% |
| | Small Town | | 217 | 12.4% |
| | City | | 684 | 39.0% |
| | Large Urban Center | | 657 | 37.5% |
| | | Total | 1754 | 100% |
| Grade(s) educators teach | Preschool age | | 189 | 10.7% |
| | 1st-3rd grade | | 437 | 24.9% |
| | 4th-6th grade | | 535 | 30.5% |
| | Middle school | | 651 | 37.1% |
| | High School | | 670 | 38.1% |
| | Adults | | 329 | 18.7% |

## 3.2. Questionnaire and Measures

The online questionnaire was released with specific guidelines for educators on how to complete it and included information on the general concepts of AIEd. It was designed to help derive conclusions about AIEd by synthesizing the potential benefits, challenges, and ethical considerations associated with its integration into educational settings. The questionnaire included questions on how AI is used by teachers during the educational process, if and how students use AI tools to support their studies, and the overall attitude of educators towards the use of AI in schools. The goal is to draw recommendations for various stakeholders and target groups such as:

- **Educators:** Teachers and instructors play the most crucial role in implementing AI technologies in the classroom and adapting their teaching methods accordingly.

- **School Administrators:** Administrators at educational institutions are responsible for making decisions regarding the adoption and implementation of AI technologies. They are concerned with improving educational outcomes, increasing efficiency, and managing resources effectively.

- **Policy Makers:** Government agencies and policymakers play a role in regulating the use of AIEd, ensuring ethical and equitable practices, and promoting access to quality education for all students and educators.

- **Parents and Guardians:** Parents and guardians are stakeholders as they are invested in the educational experiences and outcomes of their children. They are interested in how AI technologies are being used to support their children's learning and development.

- **Technology Providers:** Companies and organizations that develop AI technologies for education are stakeholders in the industry. They aim to create innovative solutions that address the needs and challenges of educators and students, while also generating revenue and staying competitive in the market.

The questionnaire comprised both quantitative (close-ended) and qualitative (open- ended) for a total of 33 questions. These included 17 multiple choice questions, two Likert, eight open-ended, and six demographic questions. The qualitative questions were initially concealed and only revealed if the educators had exhausted the provided options but still desired to include their own opinion.

The questionnaire was divided into three thematic parts. The first focused on gathering demographic information about the educators, including their experience, the grades they teach and their place of residence.        Table **1** presents the questions, measures, and frequencies for this first part of the questionnaire.

The second part of the questionnaire explored the current landscape of AIEd. Educators were asked about their integration of AI tools into their teaching practices and their individual comfort levels with AI technologies. This part also explored the perceived benefits of AI in enhancing teaching and learning experiences, and the challenges and barriers educators face. Table 2  provides a detailed overview of the questions, responses, and response frequencies from each country for this section.

Finally, the third part focused on the future of AIEd and if it potential to transform teaching and learning. Table 3 presents the questions, potential responses, and response frequencies from each participating county.

Assumptions of statistical tests were considered prior to analyses. All statistical tests were conducted using the SPSS software, version 29.0.

## 4. Results

### 4.1. Results on current landscape of AI from educators' perspective

Incorporating technology into the educational process can significantly enhance teaching and learning experiences, improving accessibility, and streamlining administrative tasks. Some methods include utilizing e-learning platforms, interactive boards, smart boards, educational apps, and software (Granić, 2022). In one recent survey (Barrett & Pack, 2023) teachers from the USA stated that they like to incorporate innovative technology in the teaching and specifically on the statement "I enjoy using new technologies for teaching" 75% of the teacher either agree or strongly agree. For educators to do so, they need to feel empowered and digital literate.

A very interesting observation arises right from the beginning of the survey, when teachers were asked to do a self-assessment of their digital competences (Clipa et al., 2023). Responses shown in Table 2, indicate that in Greece 70% consider that their digital skills are adequate, which is also the fact for most Irish teachers (68%) and Latvian teachers (60%). On the other hand, 45% of Armenian teachers believe that their skills need strengthening and 37% of Hungarian teachers believe their digital skills allow them to handle only the basics (Fig. 1).

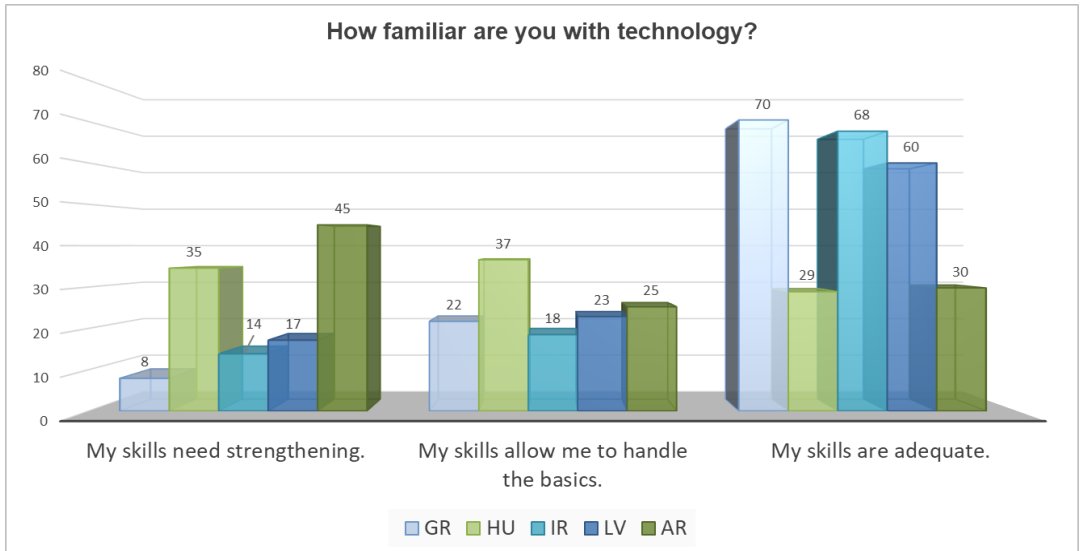

**Fig. 1** Percentage frequencies of educators' responses from Greece (GR), Hungary (HU), Ireland (IR), Latvia (LV), and Armenia (AR) to the question "How familiar are you with technology?"

Regarding the frequency of the usage of technology during the educational process (Fig. 2) most Greek respondents (43%) state that they use technology very often and only 2% state that they do not use it at all. In Ireland, the majority (53%) uses technology very often and 0% not at all. In Latvia 37% use it very often and 39% often, while in Armenia 50% states that they use technology in the educational process very often. Hungarian teachers seem to incorporate technology less. Specifically, 33% percent of teachers, state that they do not use technology at all during the educational process and 22% that they use it rarely. As shown in Fig. 2, with the exception of Hungary, in all other countries, educators use technology either often or very often in the educational process.

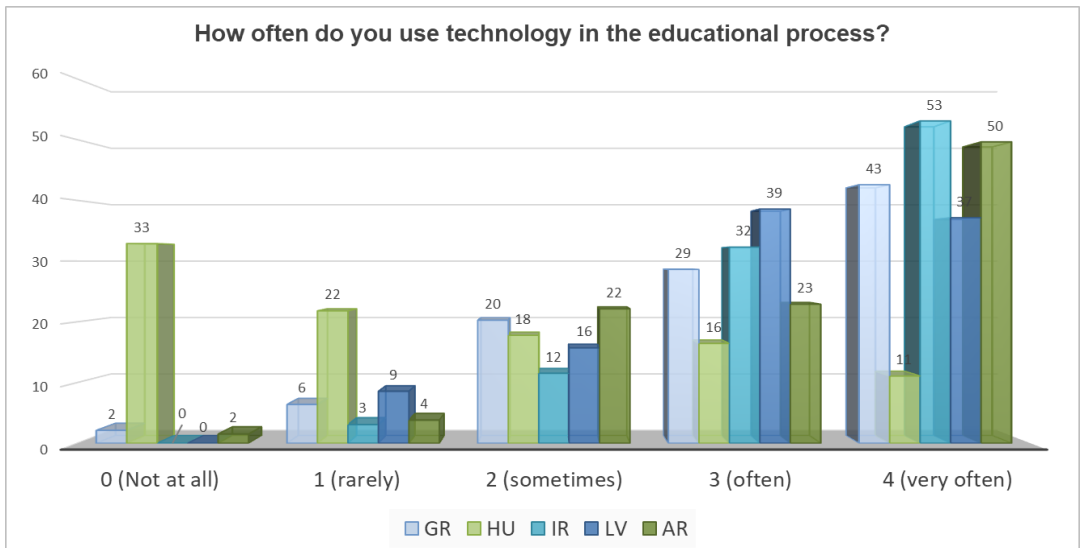

**Fig. 2** Percentage frequencies of educators' responses from Greece (GR), Hungary (HU), Ireland (IR), Latvia (LV), and Armenia (AR) to the question "How often do you use technology in the educational process?"

When it comes to AIEd most teachers from Greece, Ireland, and Armenia state that they have used AIEd tools during the educational process (63%, 55% and 50% respectively). 48% of Hungarian teachers and 61% of Latvian teachers state that they have never used tools in the educational process that include AI, as shown in Fig. 3.

While AIEd tools are increasingly being incorporated into educational settings, the level of understanding among teachers varies widely. Casal-Otero et al. (Casal-Otero et al., 2023) found that while there is growing interest in teaching AI concepts to students, many teachers themselves lack sufficient understanding of AI. Furthermore, outcomes from research paper (Holmes et al., 2022) include the recognition that most AIEd researchers are not trained to tackle the emerging ethical questions. This gap in AI literacy among educators is also arising in our survey. In Greece 67% say that they understand how the algorithms of AI methods work, but 22% says that they do not understand. The respective percentages of the other countries are 57% and 27% for Hungary, 76% and 11% for Armenia, and 76% and 5% for Latvia (Fig. 4). Interestingly, most Irish educators (53%) claim that they do not understand how AI algorithms work, and that of course is understandable, since AI literacy requires an interdisciplinary and systematic approach, and thus needs a comprehensive educators training (Walter, 2024; Luke Moorhouse, 2024).

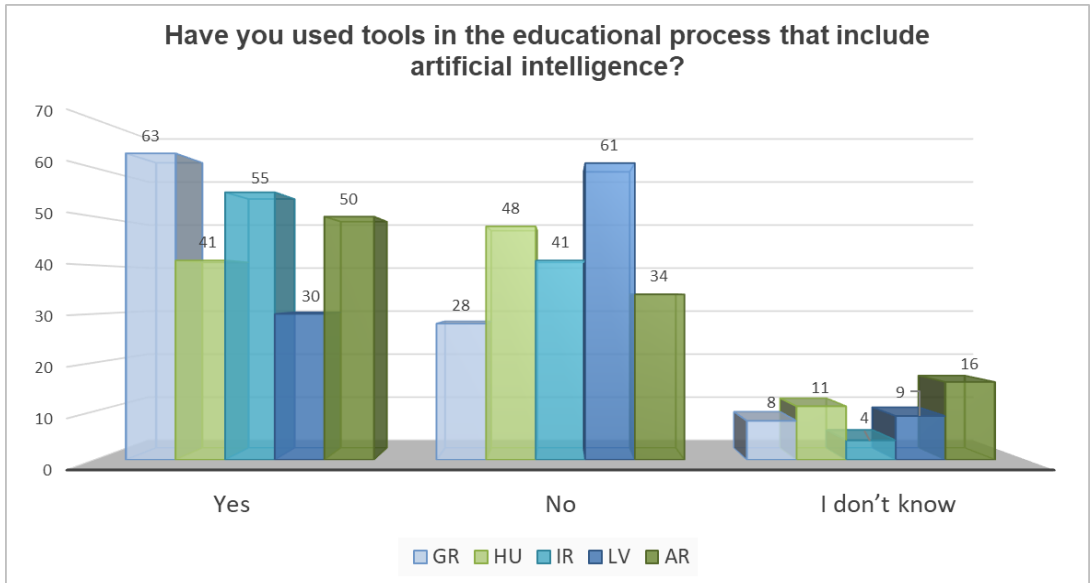

**Fig. 3** Percentage frequencies of educators' responses from Greece (GR), Hungary (HU), Ireland (IR), Latvia (LV), and Armenia (AR) to the question "Have you used tools in the educational process that include artificial intelligence?"

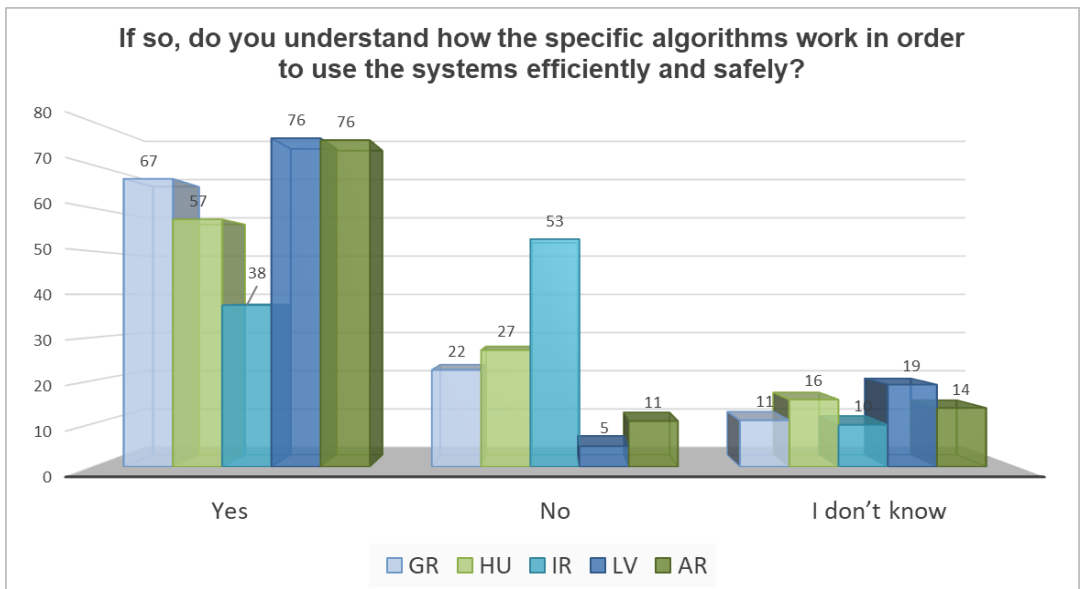

**Fig. 4** Percentage frequencies of educators' responses from Greece (GR), Hungary (HU), Ireland (IR), Latvia (LV), and Armenia (AR) to the question "If so, do you understand how the specific algorithms work in order to use the systems efficiently and safely?"

Regarding the purpose of the use of AIEd tools (Fig. 5), educators from Greece who stated that they use AIEd, report that they use them primarily for their students, namely, to capture the attention of their students (77%), to get them interested in technology (56%), and to help their student become familiar with AI (49%). In Hungary, teachers who use AIEd, do it mostly for their support and training (64%), to capture the attention of their students (60%) and to delve deeper into the lesson (53%).

As far as Ireland is concerned, they also use it mostly for their support and training (72%), but also to make the lesson understandable to all students (individualized education - vulnerable groups) (54%). In Latvia teachers use it for their own support and training (47%) but also in the class to get students interested in technology (40%). Finally, Armenian educators do also use AI tools to delve deeper into the lesson (65%), to entertain their students (60%) and to get them interested into technology (55%).

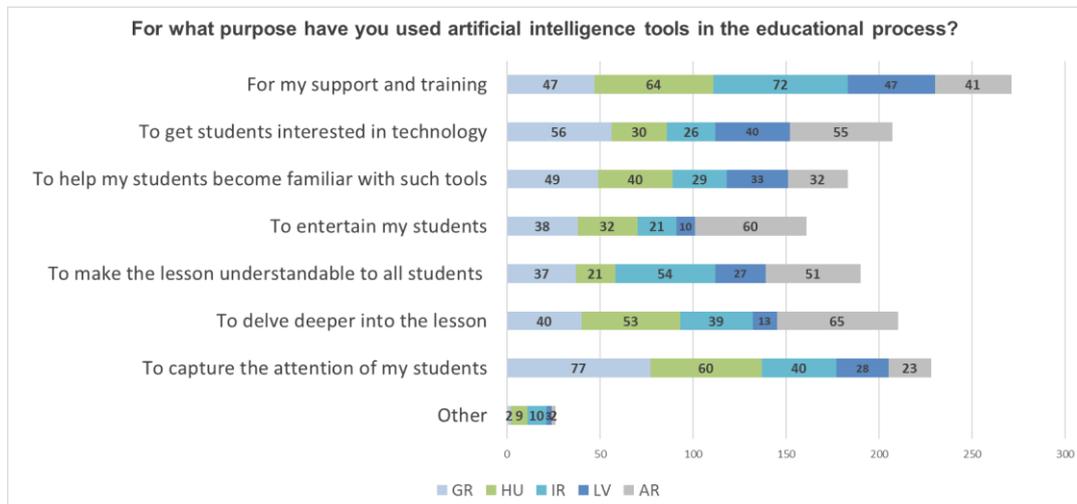

**Fig. 5** Percentage frequencies of educators' responses from Greece (GR), Hungary (HU), Ireland (IR), Latvia (LV), and Armenia (AR) to the question "For what purpose have you used artificial intelligence tools in the educational process?". The participants could choose multiple options.

Although the use of AIEd tools has been reported to be beneficial for students with learning, hearing, visual and mobility impairment (Garg & Sharma, 2020) (Rakap, 2023), form our survey we can see that only Irish educators rank it as one of their first choices, when stating the purpose of using AI.

Another noteworthy finding is that 10% of the Irish teachers state additional reasons for using AIEd. From their open-ended responses, it is evident that most of them use AI to assist them with writing texts, e.g. "To facilitate my writing as English is not my native language", "For composing emails and documents", "Writing story starters, making seating plans, writing lesson plans".

In the question "Have you noticed whether your students use artificial intelligence tools for their study?" we notice that the majority of Greek teachers (55%) answer "No" and only 15% answers "Yes". For the other countries, the responses are more balanced. In Hungary, Latvia and Armenia the majority of the educators responded with "Yes" (48%, 41% and 40%), while a significant portion of educators from Ireland responded with "No" (45%). Educators across all countries unanimously agree that students use AI tools for assistance with their academic workload. Specifically, the responses to the statement "To do their homework effortlessly" (shown in Table 2) were as follows: Greece 83%, Hungary 88%, Ireland 83%, Latvia 62%, and Armenia 53%. Another significant reason for students using AI, particularly for Ireland (35%), Latvia (38%) and Armenia (64%) is to acquire additional knowledge. Greek and Latvian teachers also noted the importance of AI "For entertainment and learning at the same time" (41% and 35% respectively).

Beyond school activities, educators also agree that students mostly use AI tools for their entertainment, but also to experiment. These are identified as the top two reasons for students' engagement with AI tools, as indicated in Table 2.

In our study, teachers highlight the benefits but also the challenges of AIEd (Florence et al., 2024; Grassini, 2023), emphasizing its potential to enhance teaching efficiency and student engagement, while they also raise concerns about potential risks (Bilstrup et al., 2020). Specifically, when asked about the potential risks associated with AI tools in education, the majority of educators declared that they are aware of these risks (Greece 53%, Hungary 56%, Ireland 73%, Latvia 49%, and Armenia 52%) (Fig. 6).

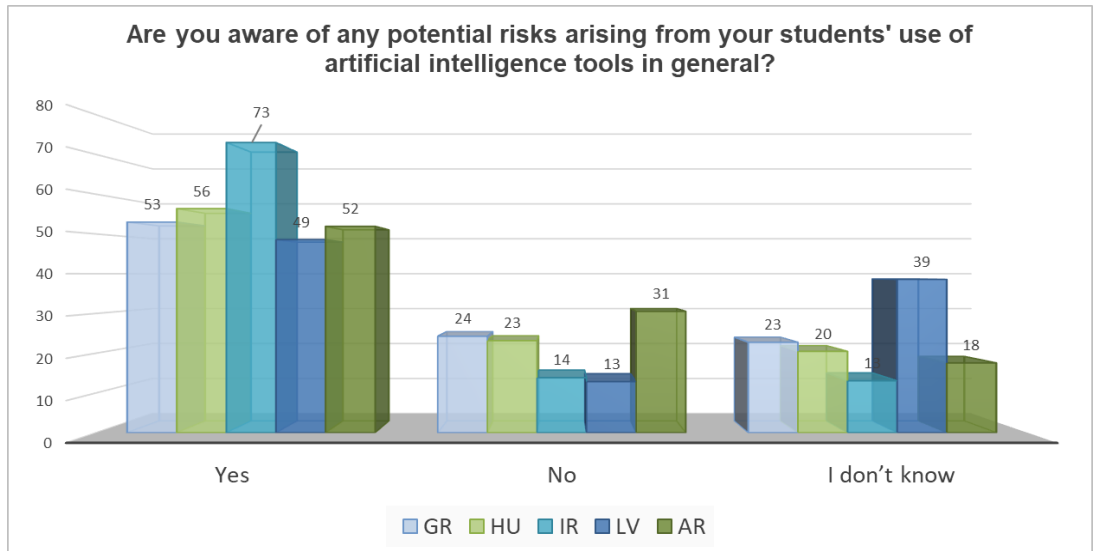

**Fig. 6** Percentage frequencies of educators' responses from Greece (GR), Hungary (HU), Ireland (IR), Latvia (LV), and Armenia (AR) to the question "Are you aware of any potential risks arising from your students' use of artificial intelligence tools in general?"

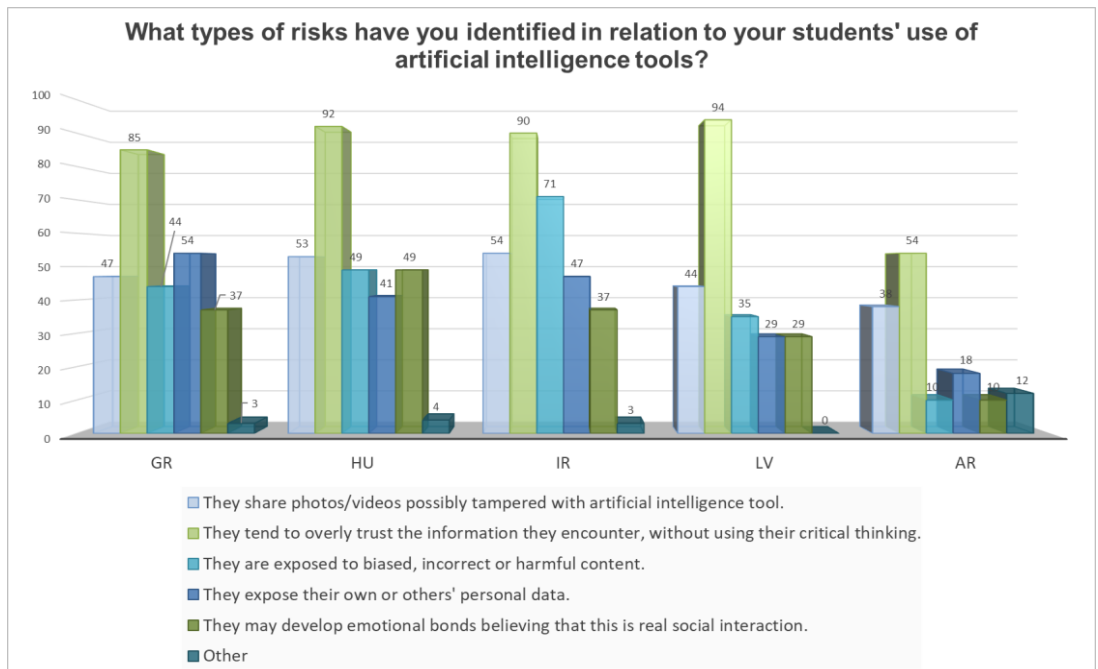

**Fig. 7** Percentage frequencies of educators' responses from Greece (GR), Hungary (HU), Ireland (IR), Latvia (LV), and Armenia (AR) to the question "What types of risks have you identified in relation to your students' use of artificial intelligence tools?". Participants were allowed to choose more than one option.

As shown in Fig. 7, they agree that the primary risk identified, is students' tendency to trust the information they find without applying critical thinking (Greece 85%, Hungary 92%, Ireland 90%, Latvia 94% and Armenia 54%). Educators from Hungary, Ireland, Latvia and Armenia also call attention to the fact that students use AI to create and then share material (e.g., photos, videos) that has been previously tampered with AI tools, which of course encloses the issue of ethical use of AI, a topic extensively discussed in previous surveys (Yusuf et al., 2024). Educators from Greece (54%) underline that their students expose their own and others' personal data when using these tools, while 71% of Irish teachers state that students are exposed to biased, incorrect or harmful content by using AI tools. Other than that, 49% of Hungarian teachers emphasize that some children might develop emotional bonds with AI robots, perceiving these interactions as real social engagements.

Furthermore, most educators across all countries state that their schools do not use AI systems for administrative work (Fig. 8).

Overall, the majority of educators who utilize these systems agree that there are mechanisms in place to ensure that personal and sensitive personal data are adequately protected.

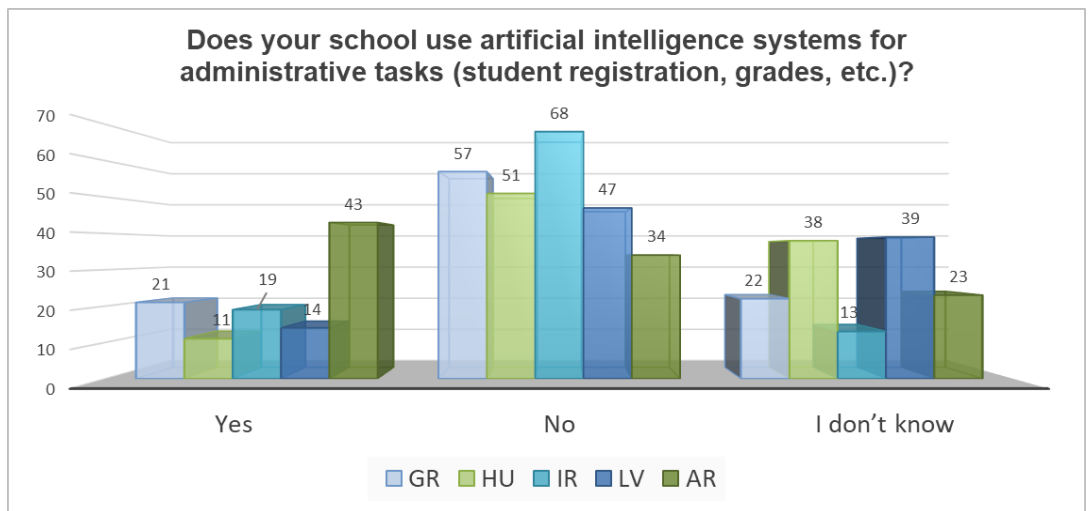

**Fig. 8** Percentage frequencies of educators' responses from Greece (GR), Hungary (HU), Ireland (IR), Latvia (LV), and Armenia (AR) to the question "Does your school use artificial intelligence systems for administrative tasks (student registration, grades, absences, etc.)?".

## 4.2. Results about the future of AI in education

Educators assert that AIEd will have a profound impact on the future of education (Table 3). According to our survey, educators unanimously believe that one of the positive effects AI will have in the future is its potential to support the work of educators. This is a common belief for 75% of Greek educators, 74% of Hungarian, 82% of Irish, 71% of Latvian educators and 64% of Armenian.

Another positive impact is its potential to facilitate personalized learning experiences for students (Chen et al., 2020). An additional finding derived from our survey, which aligns with other research findings (Ahmad et al., 2022), indicate that teachers believe AI will reduce their administrative tasks, allowing them to invest more time in teaching and mentoring students (Greece 57%, Hungary 65%, Ireland 72%, Latvia 53%, and Armenia 40%). Other ways AIEd is anticipated to positively affect the educational landscape, is by assisting in the early diagnosis of learning difficulties (Greece 41%, Hungary 35%, Ireland 50%, Latvia 26%, and Armenia 40%) and by offering great potential in educators' trainings (Greece 49%, Hungary 33%, Ireland 61%, Latvia 23%, and Armenia 42%).

It is evident from the results, that educators raise significant concerns about AI in education. Understanding public concerns about AI plays a vital role (Fast & Horvitz, 2017), as public opinion should be incorporated into the process of these regulatory decisions. Having said that, educators most important concern with the raise of AIEd, across all countries, is the failure to cultivate critical thinking (Greece 63%, Hungary 64%, Ireland 70%, Latvia 76%, and Armenia 46%) (Fig. 9).

For Armenian teachers, the biggest concern is the sharp increase in the incidents of cyberbullying and excessive online use (57%). Greek (51%) and Hungarian educators (49%) are concerned about the absence of social interactions and potential for the children to become emotionally attached to AI systems. In Ireland, 48% of educators worry about the potential exposure of children to misleading or harmful content. Additionally, 27% of Armenian teachers do also consider the risk of insufficient protection of children's personal data. The exploitation of personal content with the use of AI tools, is regarded as a lower risk by the educators (Greece & Hungary 19%, Ireland 9%, Latvia 9%, and Armenia 8%). Finally, teachers from all countries agree on the need for more guidance to enhance their knowledge and skills in using artificial intelligence tools in education (Greece 88%, Hungary 67%, Ireland 99%, Latvia 80%, and Armenia 89%).

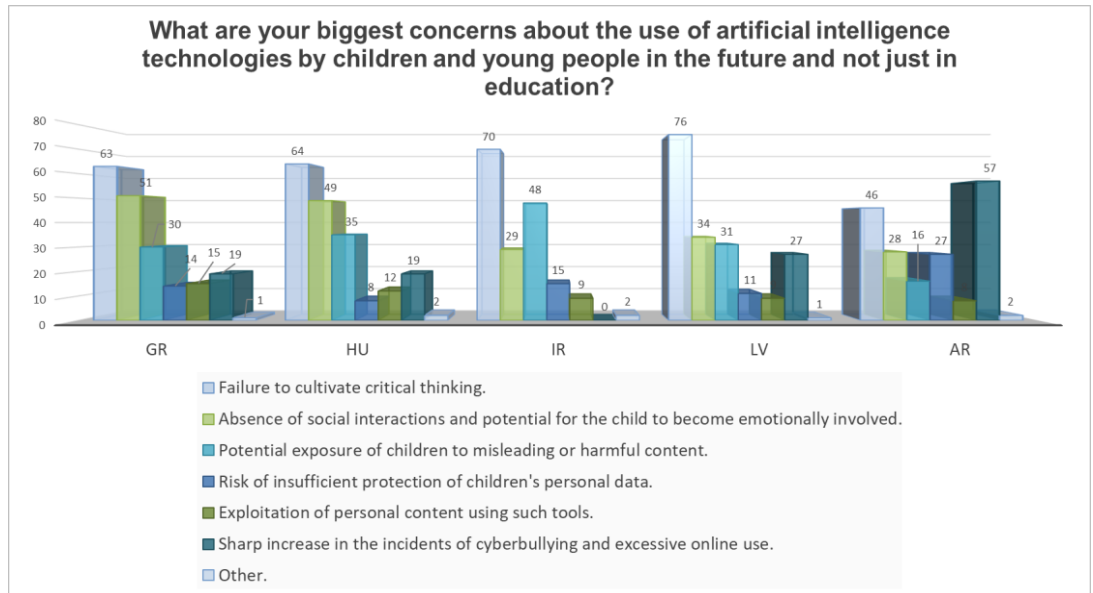

**Fig. 9** Percentage frequencies of educators' responses from Greece (GR), Hungary (HU), Ireland (IR), Latvia (LV), and Armenia (AR) to the question "What are your biggest concerns about the use of artificial intelligence technologies by children and young people in the future and not just in education?". Participants could choose multiple options.

### 4.3. Analysis on the Greek case data

In this section, we focus on the Greek data, delving deeper into the responses of Greek educators, since the authors of this paper are affiliated to the Greek Safer Internet Center of FORTH. Specifically, we perform a more detailed statistical analysis of the collected data, to unveil hidden relationships between different variables.

We use non-parametric tests, appropriate for data that do not follow a specific distribution (e.g., the normal distribution). More specifically, we use the chi-square test of independence, a method used to determine if there is a statistically significant association between two categorical variables, which allows us to conclude whether two variables are related to each other or not. The test compares the observed frequencies of each variable to the frequencies we would expect if the variables were independent. If the observed frequencies are substantially different from the expected ones, it suggests that there is an association between the variables. In summary, the chi-square test assesses whether the difference between observed and expected frequencies is due to chance or due to an association between the variables, and the extent to which the variance of one variable explains the variance of the other. In our research, we adhered to all the assumptions of the chi-square test, including that the expected frequency in each cell is five or more in at least 80% of the cells.

In addition, we use the Spearman rank-order correlation coefficient, a non-parametric measure that assesses the strength and direction of the association between two variables. The calculation of Spearman's coefficient is based on ranks rather than raw data; thus, it is appropriate for ordinal data that do not follow a specific distribution.

Based on the responses of Greek educators, gender significantly influences their behavior regarding AIEd. Specifically, a chi-square test of independence was performed to examine the relation between gender and educators' familiarity with technology. The relation between these variables was significant, $\chi_c^2$ (2, N = 1110) = 17.1, p<.001, indicating that men were more likely to be more familiar with technology than women.

The same seems to be true for the frequency of incorporating technology into the educational process. A chi-square test of independence was performed to examine the relation between gender and the frequency of using technology in the educational process. The relation between these variables was significant, $\chi^2$ (4, N = 983) = 10.65, p=.003. On the other hand, it is more likely for women to have used AI tools than men according to the chi-square test of independence ($\chi^2$ (2, N = 1110) = 12.13, p=.002).

Spearman's correlation was conducted to evaluate the relationship between age and familiarity with technology, revealing a significant inverse relationship between educators' age and familiarity with technology, $r_s([1118])$ = -.104, p <.001. The younger an educator is, the more likely it is to be more familiar with technology.

The data also reveal that it is more likely for younger age groups to have used AIEd tools during the educational process. A chi-square test of independence was performed to examine the relation between age groups and the use of tools that include artificial intelligence. The relation between these variables was significant, $\chi^2$ (6, N = 1120) = 19.96, p = .009. Specifically, it is more likely for age groups 26-35 and 36-45 to have used AIEd, than it is for age groups 46-55 and over 55. Furthermore, these latter two age groups are having higher percentages in the respond that they do not know if they have used AIEd.

A chi-square test of independence was performed to examine the relation between gender and the variable that reveals the belief that artificial intelligence will affect the educational process in the future. The chi-square test showed that the association between these variables is not significant. Moreover, variables 'age group' and 'place of residence' were tested with the variable that reveals how much artificial intelligence will affect the educational process, and again the null hypothesis was not rejected.

The null hypothesis of the chi-square test of independence was rejected for variables 'place of residence' and the biggest concern about the use of artificial intelligence technologies being the failure to cultivate critical thinking. Namely, the relation between these variables was relatively significant, ($\chi^2$ (3, N=1125) =7.5, p=.057). The statistical analysis shows that the less population the place of residence of the educator has, the less they believe that failure to cultivate critical thinking is a major risk when incorporating AI into education.

## 5. Discussion and Conclusions

AI holds tremendous potential to enhance teaching and learning in education, but its implementation must be guided by principles of equity, transparency, and ethical responsibility. By addressing challenges and ethical considerations while leveraging the benefits of AI technologies, educators and other stakeholders can work towards creating inclusive, empowering, and effective educational experiences for all.

That being said, our survey reveals that while educators from different countries hold varying views about AIEd, they do share common perspectives in certain cases. Regarding our first research question "W*hat is the current status of AIEd in the five countries we focus on? How do educators perceive it, and what are their principal areas of skepticism?*" the responses indicate that most educators from Greece, Hungary, Latvia, and Armenia claim to have a good understanding of how AI works and are aware of the potential risks. They are mostly skeptical about cultivating critical thinking in students when using AI tools, and express concerns about students will be exposed to biased data. Another interesting point is that educators mostly use AIEd for their support and training, and in the classroom to capture the attention of their students. Furthermore, educators in all countries state that their schools do not use AI systems for administrative work.

Regarding the second research question "What are the patterns, preferences, and impacts of student engagement with artificial intelligence tools in educational settings from the perspective of the educators?", findings indicate that educators believe students use AI tools in the educational context mostly for their assistance with their academic workload, stating reply "To do their homework effortlessly". Outside of school, educators believe that students predominantly use AI for their entertainment purposes.

Finally, regarding the question "*In what ways will AI shape the educational landscape from the perspective of educators?*", educators foresee that AI will highly affect the educational process in the future. They express optimism that their work will be supported by AI systems, but also that they will enable personalized learning experiences for their students. However, their primary concern for the future is that the proliferation of AI might hinder the cultivation of critical thinking skills. Last but not least, nearly all educators underline the need for assistance and more guidance to enrich their knowledge and skills in using artificial intelligence tools in education, through training seminars, workshops, but also specialized online courses.

Finally, delving deeper into the Greek data, we found that younger educators in Greece are more likely to be tech-savvy and have used AI tools. Additionally, the gender of the educators plays a role: male educators are generally more familiar with technology tools, while female educators have experimented more with AI education tools.

## Conflict of interest

No potential competing interest was reported by the authors.

## Acknowledgements

We would like to thank the Safer Internet Centers of Hungary (Király Borbála), Ireland (Jane McGarrigle), Latvia (Maija Katkovska), and Armenia (Narine Khachatryan) for their support in the collection of their national data. We would like to also thank Insafe, the European Network of Safer Internet Centers, especially Karl Hopwood, for its support in disseminating the survey.

## Funding Declaration

This study was supported by Project "SI4Kids2: The Greek Safer Internet Center SaferInternet4Kids.gr: Awareness, Helpline, Hotline", DIGITAL EUROPE, GA No. 101081266, 2022-2024.

**Table 2.** Questions and responses on artificial intelligence in education.

| Question | Answers | GR% | HU% | IR% | LV% | ARM% |
|---|---|---|---|---|---|---|
| How familiar are you with technology? | My skills are adequate | 70 | 29 | 68 | 60 | 30 |
| | My skills allow me to handle the basics | 22 | 37 | 18 | 23 | 25 |
| | My skills need strengthening | 8 | 34 | 14 | 17 | 45 |
| How often do you use technology in the educational process? | 0 (Not at all) | 2 | 33 | 0 | 0 | 2 |
| | 1 | 6 | 22 | 3 | 9 | 4 |
| | 2 | 20 | 18 | 12 | 16 | 22 |
| | 3 | 29 | 16 | 32 | 39 | 23 |
| | 4 (Very often) | 43 | 11 | 53 | 37 | 50 |
| Have you used tools in the educational process that include artificial intelligence? | Yes | 63 | 41 | 55 | 30 | 50 |
| | No | 28 | 48 | 41 | 61 | 34 |
| | I don't know | 8 | 11 | 4 | 9 | 16 |
| If so, do you understand how the specific algorithms work in order to use the systems efficiently and safely? | Yes | 67 | 57 | 38 | 76 | 76 |
| | No | 22 | 27 | 53 | 5 | 11 |
| | I don't know | 11 | 16 | 9 | 19 | 13 |
| For what purpose have you used artificial intelligence tools in the educational process? | To capture the attention of my students | 77 | 60 | 40 | 28 | 23 |
| | To delve deeper into the lesson | 40 | 53 | 39 | 13 | 65 |
| | To make the lesson understandable to all students (individualized education - vulnerable groups) | 37 | 21 | 54 | 27 | 51 |
| | To entertain my students | 38 | 32 | 21 | 10 | 60 |
| | To help my students become familiar with such tools | 49 | 40 | 29 | 33 | 32 |
| | To get students interested in technology | 56 | 30 | 26 | 40 | 55 |
| | For my support and training | 47 | 64 | 72 | 47 | 41 |
| | Other (Please specify) | 2 | 9 | 10 | 3 | 2 |
| Have you noticed whether your students use artificial intelligence tools for their study? | Yes | 15 | 48 | 35 | 41 | 40 |
| | No | 55 | 38 | 45 | 33 | 26 |
| | I don't know | 30 | 14 | 19 | 26 | 34 |
| Why do you think your students use artificial intelligence tools in their preparation and study? | To acquire additional knowledge | 21 | 29 | 35 | 38 | 64 |
| | To address their queries about the course | 25 | 38 | 20 | 20 | 38 |
| | For entertainment and learning at the same time | 41 | 25 | 9 | 35 | 28 |
| | To do their homework effortlessly | 83 | 88 | 83 | 62 | 53 |
| | Other (Please specify) | 2 | 6 | 2 | 28 | 2 |
| Have you noticed if your students use artificial intelligence tools outside their school activities? | Yes | 20 | 37 | 23 | 26 | 28 |
| | No | 33 | 28 | 42 | 19 | 8 |
| | I don't know | 47 | 35 | 35 | 55 | 64 |
| What do you believe is the reason your students use artificial intelligence tools beyond their school activities? | To acquire additional knowledge | 19 | 38 | 41 | 39 | 73 |
| | For entertainment | 85 | 80 | 69 | 61 | 54 |
| | For malicious purposes (deep fake, cyberbullying, etc.) | 10 | 8 | 14 | 11 | 5 |
| | To experiment | 67 | 67 | 55 | 61 | 49 |
| | Other (Please specify) | 13 | 2 | 0 | 6 | 5 |
| Are you aware of any potential risks arising from your students' use of artificial intelligence tools in general? (not necessarily in the educational process) | Yes | 53 | 56 | 73 | 49 | 52 |
| | No | 24 | 23 | 14 | 13 | 31 |
| | I don't know | 23 | 21 | 13 | 38 | 17 |
| What types of risks have you identified in relation to your students' use of artificial intelligence tools? | They share photos/videos possibly tampered with artificial intelligence tool. | 47 | 53 | 54 | 44 | 38 |
| | They tend to overly trust the information they encounter, without using their critical thinking. | 85 | 92 | 90 | 94 | 54 |
| | They are exposed to biased, incorrect or harmful content. | 44 | 49 | 71 | 35 | 10 |
| | They expose their own or others' personal data. | 54 | 41 | 47 | 29 | 18 |
| | They may develop emotional bonds believing that this is real social interaction. | 37 | 49 | 37 | 29 | 10 |
| | Other (Please specify) | 3 | 4 | 3 | 0 | 12 |
| Does your school use artificial intelligence systems for administrative tasks (student registration, grades, absences, etc.)? | Yes | 21 | 11 | 19 | 14 | 43 |
| | No | 57 | 51 | 68 | 47 | 34 |
| | I don't know | 22 | 38 | 13 | 39 | 23 |
| Are there mechanisms in place to ensure that personal and sensitive personal data are adequately protected? | Yes | 52 | 72 | 76 | 100 | 61 |
| | No | 9 | 0 | 0 | 0 | 11 |
| | I don't know | 39 | 28 | 24 | 0 | 28 |

**Table 3** Questions and responses on the future of artificial intelligence in education.

| Question | Answers | GR% | HU% | IR% | LV% | ARM% |
|---|---|---|---|---|---|---|
| How much do you think artificial intelligence will affect the educational process in the future? | 0 (Not at all) | 1 | 0 | 0 | 0 | 1 |
| | 1 | 2 | 3 | 0 | 1 | 2 |
| | 2 | 16 | 20 | 20 | 19 | 20 |
| | 3 | 37 | 34 | 22 | 46 | 35 |
| | 4 (Very much) | 44 | 43 | 59 | 34 | 42 |
| What positive ways, in which artificial intelligence could affect the educational process in the future, do you consider to be the most important? | Could provide personalized learning experiences for students. | 73 | 70 | 75 | 51 | 54 |
| | Could provide great potential in educators' training. | 49 | 33 | 61 | 23 | 42 |
| | Could support the work of educators. | 75 | 74 | 82 | 71 | 64 |
| | Could assist in the early diagnosis of learning difficulties. | 41 | 35 | 50 | 26 | 39 |
| | Could assist in the administrative duties of educators. | 57 | 65 | 72 | 51 | 38 |
| | Other. (Please specify). | 2 | 8 | 1 | 4 | 6 |
| What are your biggest concerns about the use of artificial intelligence technologies by children and young people in the future and not just in education? | Failure to cultivate critical thinking. | 63 | 64 | 70 | 76 | 46 |
| | Absence of social interactions and potential for the child to become emotionally involved. | 51 | 49 | 29 | 34 | 28 |
| | Potential exposure of children to misleading or harmful content. | 30 | 35 | 48 | 31 | 16 |
| | Risk of insufficient protection of children's personal data. | 14 | 8 | 15 | 11 | 27 |
| | Exploitation of personal content using such tools. | 15 | 12 | 9 | 9 | 8 |
| | Sharp increase in the incidents of cyberbullying and excessive online use. | 19 | 19 | 0 | 27 | 57 |
| | Other. (Please specify). | 1 | 2 | 2 | 1 | 2 |
| Would you like more guidance to enrich your knowledge and skills in using artificial intelligence tools in education? | Yes | 88 | 67 | 99 | 80 | 89 |
| | No | 9 | 19 | 1 | 8 | 2 |
| | I don't know | 3 | 14 | 0 | 12 | 8 |
| How could educators be more effectively trained on artificial intelligence systems? | University education. | 27 | 25 | 33 | 16 | 8 |
| | Training seminars / workshops. | 76 | 68 | 89 | 80 | 73 |
| | Specialized online courses (e.g. MOOCs). | 61 | 63 | 59 | 70 | 64 |
| | Appropriate educational material (textbooks and other teaching material). | 41 | 37 | 30 | 40 | 52 |
| | Other. (Please specify). | 2 | 4 | 2 | 1 | 53 |